\documentclass[a4paper,UKenglish]{lipics-v2018}
\usepackage[T1]{fontenc}
\usepackage[bbgreekl]{mathbbol}

\usepackage{graphics,graphicx,amsthm,amsmath,amssymb}

\usepackage{mathtools}

\usepackage{mathrsfs} 
\usepackage{float}
\usepackage[bottom]{footmisc}

\usepackage{bm}

\usepackage{url}
\usepackage{babel}

\usepackage{wasysym}

\DeclareSymbolFontAlphabet{\mathbbm}{bbold}
\DeclareSymbolFontAlphabet{\mathbb}{AMSb}

\def\({\left(}
\def\){\right)}

\def\bra#1{\mathinner{\langle{#1}|}}
\def\ket#1{\mathinner{|{#1}\rangle}}
\def\braket#1#2{\mathinner{\langle{#1}|#2 \rangle}}

\def\de{\delta\epsilon}
\def\dmin{d_{\rm min}}

\def\cVt{V_{\rm typ}}

\def\Sti{\Sigma_{*}^{\prime\prime}} 

\def\Styp{\Sigma_{\rm typ}^{\prime\prime}} 
\def\Sr{ \Sigma^{\prime}}
\def\Si{ \Sigma^{\prime\prime}}

\def\cl{{\rm cl}}
\def\pt{t_{\rm PT }}
\def\B{{B_\perp}}
\def\Bs{{B_\perp^2}}
\def\e{{\epsilon}}
\def\ve{{\varepsilon}}

\def\cP{{\mathcal P}}

\def\cO{{\mathcal O}}
\def\cE{{\mathcal E}}

\def\cV{{\mathcal V}}

\def\scP{{\mathscr P}}

\def\scH{{\mathscr H}}

\def\scS{{\mathscr S}}

\newcommand*{\medcap}{\mathbin{\scalebox{1.25}{\ensuremath{\cap}}}}%

\usepackage{microtype}

\title{Efficient population transfer via non-ergodic extended states in quantum spin glass }

\author{Kostyantyn Kechedzhi}{QuAIL and USRA, NASA Ames Research Center, Moffett Field, CA 94035, USA \\
Google Inc., Venice, CA 90291, USA}{kostyantyn@google.com}{
}
{K.K. acknowledges support by NASA Academic Mission Services, contract number NNA16BD14C.  This research is based upon work supported in part by the AFRL Information Directorate under grant F4HBKC4162G001 and the Office of the Director of National Intelligence (ODNI) and the Intelligence Advanced Research Projects Activity (IARPA), via IAA 145483. The views and conclusions contained herein are those of the authors and should not be interpreted as necessarily representing the official policies or endorsements, either expressed or implied, of ODNI, IARPA, AFRL, or the U.S. Government. The U.S. Government is authorized to reproduce and distribute reprints for Governmental  purpose notwithstanding any copyright annotation thereon.
}

\author{Vadim Smelyanskiy}{Google Inc., Venice, CA 90291, USA}{smelyan@google.com}{}{}

\author{Jarrod R. McClean}{Google Inc., Venice, CA 90291, USA}{jmcclean@google.com}{}{}

\author{Vasil S. Denchev}{Google Inc., Venice, CA 90291, USA}{denchev@google.com}{}{}

\author{Masoud Mohseni}{Google Inc., Venice, CA 90291, USA}{mohseni@google.com}{}{}

\author{Sergei Isakov}{Google Inc., Venice, CA 90291, USA}{iserge@google.com}{}{}

\author{Sergio Boixo}{Google Inc., Venice, CA 90291, USA}{boixo@google.com }{}{}

\author{Boris Altshuler}{ Physics Department, Columbia University, 538 West 120th Street, New York, New York 10027, USA}{bla@phys.columbia.edu  }{}{}

\author{Hartmut Neven}{Google Inc., Venice, CA 90291, USA}{neven@google.com }{}{}

\authorrunning{Kechedzhi et al.}

\Copyright{Kostyantyn Kechedzhi, Vadim Smelyanskiy,  Jarrod R. McClean, Vasil S. Denchev, Masoud Mohseni, Sergei Isakov,  Sergio Boixo, Hartmut Neven}

\subjclass{
}


\EventEditors{Stacey Jeffery}
\EventNoEds{1}
\EventLongTitle{13th Conference on the Theory of Quantum Computation, Communication and Cryptography (TQC 2018)}
\EventShortTitle{TQC 2018}
\EventAcronym{TQC}
\EventYear{2018}
\EventDate{July 16--18, 2018}
\EventLocation{Sydney, Australia}
\EventLogo{}
\SeriesVolume{111}
\ArticleNo{9}
\nolinenumbers 

\begin{document}

\maketitle

\begin{abstract}
Quantum tunneling has been proposed as a physical mechanism for solving binary 
optimization problems on a quantum computer because it provides an alternative to simulated annealing by directly connecting deep local minima of the energy landscape separated by large Hamming distances. However, classical simulations using Quantum Monte Carlo (QMC) were found to efficiently simulate tunneling transitions away from local minima if the tunneling is effectively dominated by a single path. We analyze a new computational role of coherent multi-qubit tunneling that gives rise to bands of non-ergodic extended (NEE) quantum states each formed by a superposition of a large number of deep local minima with similar energies. NEE provide  
a coherent pathway for population transfer (PT) between computational states with similar
energies.  In this regime, PT cannot be efficiently simulated by QMC. PT can serve as a new quantum subroutine for quantum search, quantum parallel tempering and reverse annealing optimization algorithms.  We study PT resulting from quantum evolution under a transverse field 
of an n-spin system that  encodes the energy function $E(z)$ of an optimization problem over the set of bit configurations $z$. Transverse field is rapidly switched on in the beginning of algorithm, kept constant for sufficiently long time and switched off at the end. Given an energy function 
of a binary optimization problem and an initial bit-string with atypically low energy,
PT protocol searches for other bitstrings at energies within a narrow window around the initial one.
We provide an analytical solution for PT in a simple yet nontrivial model: M randomly chosen marked bit-strings are assigned energies $E(z)$ within a narrow strip $[-n -W/2, n + W/2]$, 
while the rest of the states are assigned energy 0. The PT starts at a marked state and ends up in a superposition of L marked states inside the narrow energy window whose width is smaller than W. The best known classical algorithm for finding another marked state is the exhaustive search.
We find that the scaling of a typical PT runtime  with n and L is the same as that in the multi-target  Grover's quantum search algorithm, except for a factor that is equal to  $\exp(n /(2B^2))$ for finite transverse field $B\gg1$. Unlike the Hamiltonians used in analog quantum unstructured search  algorithms known so far, the model we consider is non-integrable and the transverse field delocalizes the marked states. As a result, our PT protocol is not exponentially sensitive in n to the
weight of the driver Hamiltonian and may be initialized with a computational basis state.  We develop the microscopic theory of PT  by constructing a down-folded dense Hamiltonian acting in the space of marked states of dimension M. It belongs to the class of preferred basis Levy matrices (PBLM) with heavy-tailed distribution of the off-diagonal matrix elements. Under certain conditions, the band of the marked states splits into minibands of non-ergodic delocalized states. We obtain an explicit form of the heavy-tailed distribution of PT times by solving cavity equations for the ensemble of down-folded Hamiltonians. We study numerically the PT subroutine as a part of quantum parallel tempering algorithm for a number of examples of binary optimization problems on fully connected graphs. 
 \end{abstract}

\section{Introduction}

Analog quantum enhanced search and optimization algorithms could, at the very least, provide a stop-gap solution for quantum applications prior to fault-tolerant universal quantum computation~\cite{preskilNISQ}. Typically the classical cost function in binary optimization problems is encoded in a $n$-qubit Hamiltonian $H_\cl = \sum_z \cE_z \ket z\! \bra z$ diagonal in the computational basis $\{\ket z\}$. The energy landscape $\{\cE_z\}$ of a typical hard optimization problem is characterized by a large number of spurious local minima. While close in energy, they can be separated by large Hamming distances. This landscape gives rise to an interesting computational primitive: given an initial bit-string with sufficiently low energy, we are to produce other bit-strings within a certain narrow range of energies $\Delta \cE_\cl$ in the vicinity of the initial state. In general, this is a hard computational task. For instance, given a solution of a SAT problem finding another solution of a similar quality is generally as hard as finding the initial solution.  Inspired by previous work in analogue quantum computing~\cite{kadowaki1998quantum,farhi2001quantum,brooke1999quantum
}, we propose the following quantum population transfer (PT) protocol: given an initial computational state $\ket{z_0}$ with classical energy $\cE_{z_0}$, we evolve with the Hamiltonian 
\begin{gather}
 H = H_\cl+H_{D},\quad H_D= - \B \sum_{j=0}^n \sigma_j^x, \label{eq:H}
\end{gather}
where $H_D$ is the driver Hamiltonian, without fine-tuning the evolution time or the strength of the transverse field $\B$. Finally we measure in the computational basis and check the energy $\cE_z$ of the outcome $\ket z$ if $z \ne z_0$. Practically, analog implementation requires rapid (diabatic) ramp on/off of the transverse field at the beginning/end of the protocol.

The distribution $\rho(E_\gamma)$ of the eigenvalues of $H$, where $H\ket{\psi_\gamma} = E_\gamma \ket {\psi_\gamma}$, is typically well localized around the mean classical energy, with an exponentially decaying tail reaching towards the low energy states. The initial state $\ket{z_0}$ is located at a deep local minimum of the classical landspace $\{\cE_z\}$, at the tail the distribution $\rho(E_\gamma)$. 
The non-diagonal matrix elements $-\B$ give rise to hopping between states separated by one bit-flip. Matrices with diagonal disorder and hoping between neighbors correspond to the Anderson model introduced in the context of transport and localization in disordered media~\cite{anderson1958absence}. 
The transverse field $\B$ couples the local minima via perturbation theory in a high order given by the Hamming distance between them. 
In this model (\ref{eq:H}), as well as  in the original Anderson model,  there exist bands of localized and extended states separated in energy by a so-called 
``mobility edge''. It  was demonstrated in  Ref.~\cite{altshuler2010anderson}  that localization is responsible for the failure of  Quantum Annealing to find a solution of  the  constraint satisfaction problem (although,  the detailed analysis of this effect is still lacking~\cite{knysh2010relevance,knysh2016zero}).
Nonetheless, extended states could provide a mechanism for population transfer away from the initial state. 
In spin-systems with transverse field the existence of a so-called ``mobility edge'' at the tail of the distribution $\rho(E_\gamma)$, separating in the energy spectrum localized and delocalized eigenstates of $H$, has been recently studied in Refs.~\cite{smelyanskiy2002dynamics,laumann2014many,mossi2017many}.

The population transfer corresponds to the formation of a ``conduction band'' that could be understood from the following arguments. We express the probability of a transition from $\ket {z_0}$ to $\ket z$,
\begin{align}
  P(t,z|z_0) = \left|\sum_\gamma \braket z {\psi_\gamma} \braket{\psi_\gamma} {z_0} e^{-i E_\gamma t} \right |^2\;,\label{eq:ptz}
\end{align} 
in terms of the eigenstates $\{\ket{\psi_\gamma}\}$ and eigenvalues $\{E_\gamma\}$ of the system Hamiltonian $H$. In the delocalized phase the state $\ket{z_0}$ has a sizable overlap with a large set of eigenstates of size $\Omega$ with energies within some range $|E_\gamma - E_{\gamma'}| \sim \Delta E$. These are the eigenstates that dominate the sum in Eq.~\eqref{eq:ptz}. The eigenstates in this set  posses an important property extensively studied in the theory of transport in disordered systems~\cite{Kravtzov2015RzPr,altshuler2016multifractal,altshuler2016nonergodic}. They have largely overlapping supports over a support set $\scS$ of bit-strings. This implies, from Eq.~\eqref{eq:ptz},  that after a typical population transfer time $\pt \propto 1/\Delta E$ and for any initial state $z_0 \in \scS$ the population is spread over the entire set $\scS$, that is $P(\pt, z_1|z_0) \sim 1/|\scS|$ for all $z_1 \in \scS$. The conduction band is formed by the eigenstates within a spectrum width $\Delta E$ associated with the bit-strings in $\scS$. From the point of view of condensed matter physics, the eigenstates that overlap with $\ket{z_0}$ are non-ergodic. Nevertheless, they form a conduction miniband with energies below the quantum spin glass transition at the tail of $\rho(E_\gamma)$~\cite{laumann2014many,altshuler2016nonergodic}. 

The formation of the conduction band explained above is the physical mechanism that we intend to exploit in the PT protocol to solve the computational primitive defined above. It is well established that simulating unitary time quantum dynamics in the delocalized phase to approximate $P(\pt, z_1|z_0)$ can not be done efficiently by known numerical techniques, such as quantum Monte-Carlo or tensor network methods, due to the coherent many-body nature of this transport phenomena. 
 In addition to tunneling the transverse field $\B$ gives rise to shifts in the classical energies $\cE_\cl$ distributed over the width $\Delta \cE_\cl$. This limits how narrow the target window of classical energies in the primitive can be. 

More generally, the PT protocol can provide a useful primitive to explore energy landscapes on the way to lower energy states for optimization, reverse annealing~\cite{LidarNishimoriReverse} and quantum machine learning~\cite{carleoTroyer2017,biamonte2017quantum}. The  output of PT  $z$ can be used as an input  of a classical optimization heuristic such as simulated annealing  or parallel tempering in a ``hybrid'' optimization algorithm~\cite{Neven2016} where quantum and classical steps  can be used sequentially to gain the complementary
advantages of both~\cite{chancellor2017modernizing}. A quantum advantage of the population transfer primitive would imply an advantage of such quantum parallel tempering over similar classical algorithms. 

We propose a theoretical approach to analyze this problem with detailed analysis presented in~\cite{IB}.  Here we provide the results. Our approach exploits the existence of two relevant energy scales. The first scale is the width of the non-ergodic conduction miniband $\Delta E$. The second scale is the typical change in classical energy $\cE_\cl$ corresponding to one spin flip. Because of the large Hamming distances separating states in the support of the conduction band $\scS$, the effecitive coupling elements that couple them correspond to high order in perturbation theory in $\B$, and therefore $\left|\cE_\cl \right| \gg \Delta E$. The dynamics within the miniband is described by the effective downfolded Hamiltonian
\begin{align}
  \scH = \sum_{j=1}^M \ve_j \ket j \bra j + \sum_{j,k=1}^M V_{jk} \ket j \bra k\;. \label{eq:Heff}
\end{align}
The sum is over a size $M$ of the subset of computational basis states $\ket {z_j}$ sufficiently wide such that it contains the support set $\scS$. The $\ve_j$'s are appropriately renormalized energies of the Hamiltonian $H_\cl$. The non-diagonal matrix elements $V_{jk}$ correspond to the sum over all elementary spin-flip processes that begin in state $\ket j$, proceed through virtual states separated by energies at least $\cE$ from the miniband, and return back to the miniband only at the last step, at the state $\ket k$. We emphasize that in general $V_{jk}$ takes into account all loops where the process returns back to the same virtual state without visiting the miniband. 

In this paper we apply this analytical framework to "impurity band" model which demonstrates a quantum spin glass behavior yet allows analytical description of the quantum dynamics in the course of the PT protocol. In the second part of the paper  we present numerical analysis of a set of more practical models defined by 2-local Hamiltonians. The impurity band model is defined by the Hamiltonian, 
\begin{align}
 H = H_\cl - \B \sum_{j=0}^n \sigma_j^x,\quad  H_\cl = \sum_{j=1}^M \cE(z_j) \ket {z_j} \bra{z_j}\label{Hcl2}
\end{align}
where the $n$-bit-strings $\{z_j\}_{j=1}^M$ are chosen uniformly at random from all bit-strings of length $n$, there are $M \gg 1$ marked states $\ket{z_i}$, with energies $\cE_{z_j} = -n + \ve_j$. The $\varepsilon_j$'s are independently distributed around $0$ with a narrow width $W \ll 1$ to be discussed below. All other states have energy $0$ and are separated by a large gap $\sim n$ from the very narrow band of marked states. The typical distance between marked states is $n/2$. If  $M$ is exponentially  large in $n$  the  typical distance  $d_{\rm min}$ to the nearest marked  state is  much smaller than  $n/2$ but remains extensive $\dmin={\cal O}(n)$. As such, each marked state $\ket{z_j}$  is a deep local minimum of $\cE(z)$ coupled to other marked states via transverse field induced multiqubit tunneling with amplitude decreasing exponentially with Hamming distance $d$.

We obtain an explicit analytical form for the statistical properties of the PT dynamics in the above model (\ref{Hcl2}) by deriving in the form of Eq.~(\ref{eq:Heff}) an effective down-folded Hamiltonian in the energy strip associated with the PT~\cite{IB}
\begin{equation}
\scH_{ij}=\delta_{ij}\epsilon_j+(1-\delta_{ij})\cV_{ij} \sqrt{2} \sin \phi(d_{ij})\;.\label{eq:IBH}
\end{equation} 
Here the diagonal elements $\epsilon_j$ are given by the marked state energies counted off from the center of the impurity band shifted due to the effect of the transverse field $\sim \Bs$. Their PDF  is assumed to be exponentially bounded with some width~$W$. 

Explicit analytical form of the off-diagonal elements is obtained using WKB approach. In Eq.~(\ref{eq:IBH}) $\phi(d)\equiv \phi(E^{(0)},d)$ is a WKB phase that describes the oscillation of the matrix elements with the Hamming distance. The tunneling amplitude $\cV_{ij}$ equals 
\begin{equation}
\cV_{ij} \equiv V(d_{ij}),\quad V(d)=\sqrt{A(d/n, \B)}\,\,\frac{ n^{5/4} \,  e^{-n\theta(\B)} }{\sqrt{\binom{n}{d}}}\;,\label{eq:V2}
\end{equation}
 where $i$ $\neq$$ j$ and the coefficient $A(\rho, \B)$ is a smooth function of its arguments. The function $\theta(\B)$ is given in~\cite{IB}. Below we use its asymptotical form in the limit  $\B\gg1$, 
\begin{equation}
\theta \simeq \frac{1}{4 B_{\perp}^{2} }+\frac{1}{24 B_{\perp}^{4}}+\frac{1}{60B_{\perp}^{6}}+\ldots.\; \label{eq:g-exp}
\end{equation}
In this limit $\theta\ll 1$. 
We shall refer to $\scH$ in (\ref{eq:IBH}) as the Impurity Band (IB) Hamiltonian.

 The typical matrix element corresponds to tunneling at distance $n/2$ given by, 
 \begin{gather}
 V_{\rm typ}\sim n^2 2^{-n/2}e^{-n/(4\Bs)}.\label{eq:typ}
 \end{gather}
 At the same time the typical distance from a  marked state to its nearest neighbor is extensive $d_{\rm min}=\cO(n)$  which corresponds to a matrix element, see Eq.~(\ref{eq:V2}), exponentially larger than the  typical value $V_{\rm typ}$. Therefore there is a hierarchy of off-diagonal matrix elements of $\scH_{ij}$. The off diagonal matrix elements of random realizations of $\scH$ are described by a heavy-tailed probability density function~\cite{cizeau1994theory,monthus2016localization}.  Such random matrices are called Levi matrices.

The  PDF of the rescaled squared amplitudes $w_{ij}=V^2(d_{ij})/V_{\rm typ}^{2}$ can be obtained in the explicit form~\cite{IB},
\begin{equation}
{\rm PDF}(w)=\frac{1}{w^2\sqrt{\pi \log w}},\quad w\in[1,\infty). \label{eq:poly}
\end{equation}
The particular form of scaling is the direct consequence of the fact that our problem has no "structure":   the tunneling matrix elements depend only on Hamming distance and marked states are chosen at random. 

The key difference  of the ensemble of matrices $\scH_{ij}$ from  Levi matrices studied in the literature \cite{tarquini2016level,monthus2016localization,metz2010localization,cizeau1994theory} is that the dispersion,  $W$, of the diagonal matrix elements is much larger than the typical magnitude of the off-diagonal elements  $V_{\rm typ}$. Therefore $\scH_{ij}$ can be called
preferred basis Levi matrices (PBLM). 

We note that the existence of heavy tails in the PDF of the off-diagonal matrix elements of the down-folded  Hamiltonian $\scH$ is due to  the infinite dimension of the Hilbert space of the original problem (\ref{eq:H}) for  $n\rightarrow \infty$. This happens because the exponential decay of the matrix elements with the Hamming distance $d$ is compensated by the exponential growth of the number of  states at the  distance $d$  from a given state. We expect that this PBLM structure is a generic feature of the effective Hamiltonians for PT at the tail of  the density of states in quantum spin glass problems. 

The competition between the exponential decrease of the matrix element and the increase of the number of neighbors at distance $d$ can result in eigenstates $\ket{\psi_\beta}$ of $H$ associated  with the impurity band becoming delocalized over a large subset  of marked states $\scS_\beta$  with size $1\ll |\scS_\beta|\propto M^\alpha$ and  $0<\alpha \leq 1$.  For $\alpha=0$ the eigenstate $\ket{\psi_\beta}$ is localized, for $\alpha=1$ the eigenstate is delocalized in the  entire space of  marked states. For $0<\alpha<1$ the eigenstate can be considered "non-ergodic" and its support  set $\scS_\beta$ is sparse in the space of the marked states. The PBLM matrices support non-ergodic delocalized states when the width $W$ is much bigger than the largest off-diagonal matrix element in a typical row of $\scH_{ij}$ and much smaller than the largest off-diagonal element in a matrix 
\begin{equation}
V_{\rm typ} M^{1/2}\ll W\ll V_{\rm typ} M\;.\label{eq:VMV}
\end{equation}
For smaller dispersion $W\apprle V_{\rm typ} M^{1/2}$ the matrix eigenstates  are ergodic while for $W \apprge  V_{\rm typ} M$ the eigenstates are localized. This non-ergodic regime is a distinct feature of the PBLM and is absent in Levi matrices. Such phase diagram resembles the one in the Rosenzweig-Porter (RP)  model~\cite{Kravtzov2015RzPr,facoetti2016non}. The difference of RP from PBLM is that  the statistics of the off-diagonal matrix elements in the RP ensemble are Gaussian \cite{rosenzweig1960repulsion} rather than polynomial (\ref{eq:poly}). In this paper we focus on PT transfer within the non-ergodic delocalized phase, which is more likely to generalize to other models. We note that the localized phase does not support population transfer.

In the delocalized phase eigenstates with largely overlapping supports $\medcap_{\beta}\scS_\beta$\;$\approx$\;$\scS(z_j)$ form narrow mini-bands.  The mini-band width $\Gamma$ may be interpreted as the inverse scrambling time and determines the width of the  plateau in the Fourier-transform of the typical transition probability $\tilde P(\omega,z|z_j)$ ~\cite{Kravtzov2015RzPr}.\footnote{The same plateau width  characterizes the frequency dependence of the  eigenfunction  overlap correlation coefficient $K(\omega)=M\sum_{j=1}^{M}\sum_{\beta,\beta'}   |\braket{j}{\psi_\beta}|^2   |\braket{j}{\psi_{\beta'}}|^2 \delta(\omega-E_{\beta}+E_{\beta'})$~\cite{Kravtzov2015RzPr}.}
 In other words, the  significant PT of $P(t,z|z_j)$   from the initial marked state $\ket{z_j}\in \scS $ into other states of the same miniband $\scS$ occurs over time $\pt\sim 1/\Gamma$. 

Because of the PBLM structure of the Hamiltonian $\scH$ one can expect that the runtime of the PT protocol $\pt$ will have  a heavy-tailed PDF  whose form is of practical interest. It is closely related to the PDF of the miniband widths  $\Gamma\sim 1/\pt$. We obtained the PDF$(\Gamma)$  using the cavity method for random symmetric matrices \cite{abou1973selfconsistent,cizeau1994theory,burda2007free,tarquini2016level}. 

\begin{figure}[ht]
  \includegraphics[width= 3.35in]{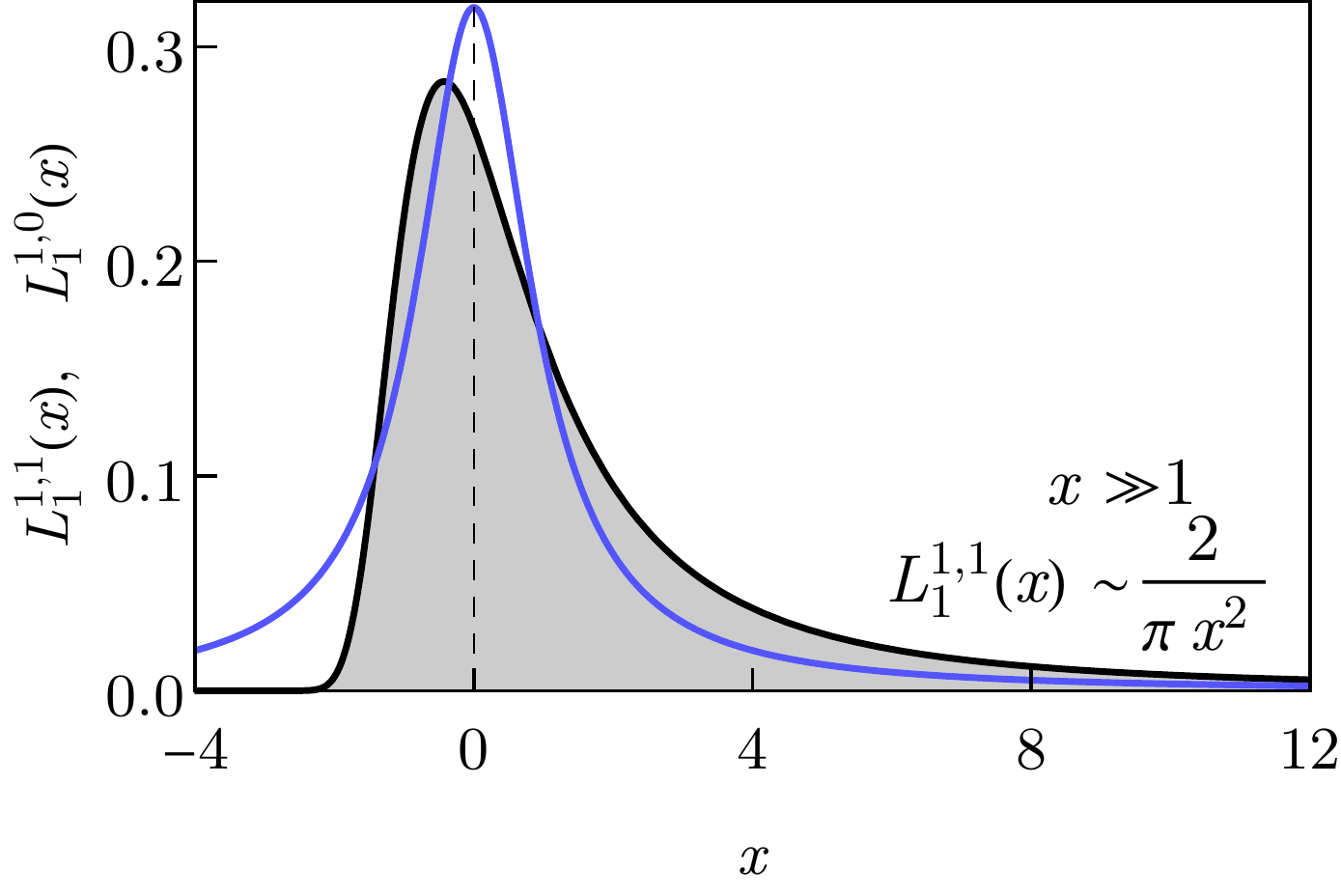}
  \caption{Black solid line shows the plot of the  Levi alpha-stable distribution  $L_{\alpha}^{C,\beta}(x)$  \cite{cizeau1994theory}   with tail index $\alpha=1$, asymmetry  parameter $\beta=1$  and  unit scale parameter $C=1$.  Inset shows asymptotic behavior of the distribution at large positive $x$.  At  $-x \gg 1$ the function decays steeply as a double exponential,  $\log L_{1}^{1,1}(x)\propto -
 \exp(-\frac{\pi}{2} x)$.  Blue line shows the Cauchy distribution $L_{1}^{1,0}(x)=\frac{1}{\pi(1+x^2)}$. 
 We follow here the definition introduced in   \cite{cizeau1994theory} and used in subsequent papers on Levi matrices in physics literature. In mathematical literature  \cite{
voit2013statistical} a different definition  is usually used, corresponding to $f(x; \alpha, \beta,C^{1/\alpha},0)=L_{\alpha}^{C,\beta}(x)$. }\label{fig:st-dist}
\end{figure}

Previously cavity equations were solved only in their linearized form, i.e.,  near the localization transition. We were able to solve  the fully nonlinear cavity equations in the delocalized non-ergodic phase~\cite{IB}. We obtained boundaries of the non-ergodic phase analytically in terms of the ratio of $W/V_{\rm typ}$ and the form of  the PDF $\scP(\Gamma)$ inside the phase. 
It is given by the alpha-stable Levi distribution \cite{gnedenko1954limit,cizeau1994theory} with the tail index 1, see Fig.~\ref{fig:st-dist}
\begin{gather}
\scP(\Si)=\frac{1}{C} L_{1}^{1,1}\(\frac{\Si-\Styp}{C}\)\;,\label{eq:scPIm} \\
\Styp=\mu_\Omega\Sti,\quad C=\sigma_\Omega \Sti\;.\label{eq:SC}
\end{gather}
Here $\Styp$ is a   shift  of the distribution and $C$ its scale parameter  (characteristic width) and we introduced a notation  $\Sigma_\ast^{\prime\prime}=\pi V^2_{\rm typ}/(W/M)$.  
\begin{align}
\mu_\Omega & \simeq \frac{1}{\sigma_\Omega}+\frac{2\sigma_\Omega(1-\gamma_{\rm Euler})}{\pi}\;.\label{eq:mu-Omega}\\
\sigma_\Omega &=\sqrt{\frac{\pi}{4\log \Omega}}\;.\label{eq:sigmaOmega0}
\end{align}
Here $\Omega$ is the number of states in the miniband. 
This number $\Omega=(\pi M V_{\rm typ}/W)^2$ is a square function of the ratio of the typical tunneling matrix element $V_{\rm typ}$ to the level separation $W/M$.

We introduce the scaling of the width of the distribution of $\epsilon_m$ with the matrix size $M$,
\begin{gather}
W= \lambda M^{\gamma/2} V_{\textrm{typ}}\,, \label{eq:param}
\end{gather}
where $\gamma$ is a real non-negative parameter that controls the scaling of the ratio of the  typical diagonal to off-diagonal matrix element $V_{\textrm{typ}}$ given in Eq.~(\ref{eq:typ}), and $\lambda$ is an auxiliary constant of order one. With this scaling ansatz we get~\cite{IB}, 
\begin{equation}
\Omega=\(\frac{\pi}{\lambda}\)^2 M^{2-\gamma}\;.\label{eq:Omegac}
\end{equation}
 
Using the above scaling ansatz (\ref{eq:param}) and  the expressions for  $\sigma_\Omega$ (\ref{eq:sigmaOmega0}) and $\mu_\Omega$ (\ref{eq:mu-Omega}) we obtain,
\begin{align}
\Styp  &\simeq \frac{2 \pi^{1/2} }{\lambda}\, \cVt M^{1-\gamma/2} (\log \Omega)^{1/2}\;,\label{eq:shift}\\
C& \simeq \frac{\pi^{3/2}}{2\lambda}\,\,\cVt M^{1-\gamma/2}   (\log \Omega)^{-1/2}\;.\label{eq:width}
\end{align}
The most probable  value of the miniband width is $\Gamma_{\rm typ}= V_{\rm typ}(\pi\Omega \log\Omega/4)^{1/2}$, and its characteristic dispersion 
$\pi \Gamma_{\rm typ}/(4\log\Omega)$. 
In a non-ergodic delocalized phase $M\gg \Omega\gg1$ and  the typical PT time $\pt\sim1/\Gamma_{\rm typ}$ obeys the condition
\begin{equation}
(M\log M)^{-1/2}\ll \pt V_{\rm typ}\sim (\Omega\log\Omega)^{-1/2} \ll 1\;.\label{eq:VMVOmega}
\end{equation}

\section{\label{sec:compl} Complexity of the Population Transfer protocol}

Starting at $t=0$ from a marked state $\ket{z_j}$ the probability for the population to be transferred to other marked states is $1-\psi^{2}(z_j, t)$. 
At the initial stage 
the survival probability $\psi^{2}(z_j, t)$ decays exponentially with the mean decay time $1/\Gamma_{j}=1/(2\Si_j)$.

The  initial marked state $\ket{z_j}$ decays into the eigenstates $\ket{\psi_\beta}$ of the IB Hamiltonian $\scH$ with typical energies $E_\beta$ inside the narrow interval  corresponding to the miniband associated with $\ket{z_j}$. It has width  $\Si_j$ and is centered around $\scH_{jj}=\e_j$.  Typical classical energies $\e$ of the bit-strings measured at the end of PT protocol will obey the probability distribution $\cP(\e-\e_j-\Sr_j)$ where $\Sr_j$ is the self-energy shift of the marked state $\e_j$ and $\cP$ is the rescaled Cauchy distribution shown in Fig.~\ref{fig:st-dist} which reads~\cite{IB},
\begin{gather}
\cP(\Sigma^\prime)=\frac {1}{\pi} \frac{\Sigma^\prime_{\rm typ}}{\left(\Sigma^\prime_{\rm typ}\right)^2+\left(\Sigma^\prime\right)^2}, \quad \Sigma^\prime_{\rm typ} = \Sigma_\ast^{\prime\prime}\sqrt{\frac{4\log M}{\pi}}.
\end{gather}

 The  success of PT protocol  is  to find a bit-string distinct from $z_j$ at a time $t$ with energy inside the window  $\Delta \cE_{\rm cl}$ around $\e_j$.  The  PT success time therefore equals
\begin{equation}
t_{\rm PT}^{j}= \frac{1}{2\Si_j p_{\Delta\cE}},\quad p_{\Delta\cE}=\int_{0}^{\Delta\cE_{\rm cl}}\cP\(\e-\Sr_j-\frac{\Delta\cE_{\rm cl}}{2} \)d\e\;.\nonumber\label{eq:ptj}
\end{equation}
Here $p_{\Delta\cE}$ is the probability of detecting a bit-string inside the target window  $\Delta \cE_{\rm cl}$ under the condition that initial state has decayed.
Assume that the PT window is as wide as the typical miniband width, $\Delta \cE_{\rm cl}=\Styp$.  In this case $p_{\Delta\cE}$ differs from 1 only by a constant factor that does not depend on $M$. Therefore after a sufficiently long time a solution, a bit-string inside the PT window, is detected with finite  probability. Because the initial state  $\ket{z_j}$ is picked at random the typical  time  to success  of PT   $\pt\sim 1/ \Styp$ corresponds to the inverse typical  width of the miniband.   All of the states in a miniband are populated at (roughly) the same time $\pt$ because  the  transition rate to a subset of states on a distance $d$ away from $\ket{z_j}$ depends on $d$ very weakly. This is a result of a cancellation between the combinatorial number $\binom{n}{d}$ of states (and hence decay channels) at distance $d$ from a given marked state and the dependence of the matrix element squared on $d$, see (\ref{eq:V2}).  
 
 We characterize the PT  by the relation between the typical success time of PT  $\pt$ and the number of states $\Omega$ over which the population is spread  during PT
\begin{equation}
\pt\sim\frac{1}{\cVt\,\sqrt{\Omega \log\Omega}}\sim \left(\frac{2^{n}}{n \Omega \log\Omega} \right)^{1/2} e^{2\theta n}\;,\label{eq:Gr-rel}
\end{equation}
where we set $\Delta\cE_{\rm cl}\sim \Sti$ (see discussion above).  The time $t_{\rm G}$ for the Grover algorithm for unstructured quantum search to find $\Omega$ items in a database of the size $2^n$ is $t_{\rm G}\sim (2^{n}/\Omega)^{1/2}$. 
PT time $\pt$ scales worse than Grover time $t_{\rm G}$ by an additional exponential factor $e^{2\theta n}\simeq e^{ \frac{n}{2B_{\perp}^{2}}}$   (\ref{eq:g-exp}). At large transverse  fields $1\ll \B=\cO(n^0)$ the scaling exponent is small $2 \theta \ll 1$.

\section{\label{sec:Grover} Comparison with the analogue  Grover search}

Inspired by the Hamiltonian version of Grover algorithm proposed in \cite{farhi1998analog} we consider the PT protocol in the IB model $H_{\rm cl}$ starting from the  ground state of $H_D$   which is a
fully symmetric state $\ket{S}=2^{-n/2}\sum_{j=1}^{n}\ket{z}$  in a computational basis. This protocol can be implemented   by adjusting  the value of transverse field $\B$ $\approx$ 1 so that  the ground state energy of the driver is set near the center of the IB. Then we can replace the full driver with the projector on its ground state, $H_D\rightarrow -n\B\ket{S}\bra{S}$. 
The quantum evolution is described by the  Hamiltonian
\begin{equation}
H_{\rm G}=-n\B \ket{S}\bra{S}+\sum_{j=1}^{M}\cE(z_j)\ket{z_j}\bra{z_j}\;,\label{eq:HG}
\end{equation}
with the initial condition  $\ket{\psi(0)}=\ket{S}$.  In the case where all impurity energies are equal to each other, $\{\cE(z_j)=-n\}_{j=1}^{M}$, and $\B=1$ the Hamiltonian $H_{\rm G}$  is a generalization of the analog version of Grover search \cite{farhi1998analog} for the case of  $M$ target states. The system performs Rabi oscillations between the initial state $\ket{S}$ and  the state which is an equal superposition of all marked (solution)  states. Time to solution is the half-period of the oscillations, the "Grover time" $t_{\rm G}$
\begin{equation}
t_{\rm G}=\frac{\pi}{2n\B} \sqrt{ \frac{2^n}{M}}\;.\label{eq:Grover}
\end{equation}
Hamiltonian versions of Grover search with transverse field driver whose ground state were tuned at resonance with that of the solution state were considered in~\cite{farhi2000quantum,childs2002quantum}.

We assume as before that  marked state energies   take distinct values $\cE(z_j)=-n+\e_j$ randomly   distributed over some  narrow  range  $W$.
We investigate the effect of systematic error in the Grover diffusion operator~\cite{grover1997quantum}. In the Hamiltonian formulation \cite{farhi1998analog}
this corresponds to the deviation from unity of the parameter $\B$  that controls the weight of the driver in (\ref{eq:Grover}). We will define the driver error $\e_0$ by,
\begin{equation}
\B=1-\frac{\e_0}{n}\;.\label{eq:e0}
\end{equation}

Assuming that $N\gg M$ one can  instead of the state $\ket{S}$ consider the  decay of the  state $\ket{0} \equiv \frac{1}{\sqrt{N-M}}\sum_{j=M+1}^{N}\ket{z_j}$, where $j\in[1,M]$ corresponds to marked states. 
We use (\ref{eq:e0}) and omit constant terms and small corrections $\cO(M/N)$ in $H_G$.  The  non-zero matrix elements  $H_{G}^{ij}=\bra{i}H_G\ket{j}$ 
in this subspace $\scS$ have the form 
\begin{equation}
H_{G}^{jj}=\e_j,\quad  H_{G}^{j0}= -(1-\delta_{j0})V,\quad V=n 2^{-n/2}\;,\label{eq:HijG}
\end{equation}
where  $j\in[0,M]$ and $H_{G}^{j0}=H_{G}^{0j}$.
On a time scale $t\ll1/\de=M/W$ much smaller than the inverse spacing of the  energies $\e_j$
the quantum evolution with  initial condition $\ket{\psi(0)}=\ket{0}$ corresponds to the decay of the discrete state with energy $\e_0$ into the  continuum \cite{mahan2013many} with the finite spectral width $W$ \cite{kogan2006analytic}.
We assume that 
$\e_0\gg W$ while the spread of the marked state energies
 $W\lesssim V\sqrt{M}$,
so that absent driver errors, PT time  would follow a Grover-like scaling law $t\sim1/(V \sqrt{M})$.

The state $\ket{0}$ is coupled non-resonantly to a continuum with narrow bandwidth.  The   expression for the population transfer  to the marked states can be obtained  from the time-dependent perturbation theory in the parameter $\e_0/W$
\begin{equation}
\sum_{m=1}^{M}|\psi_{m}(t)|^2=\frac{2 M V^2}{\e_0^2}\(1-\cos(\e_0 t)\frac{\sin( Wt/2)}{W t/2}\right)\;.\nonumber
\end{equation}
Maximum transfer occurs at the time $t_0=\pi/\e_0$ with the total transferred probability $p_0=4M V^2/\e_0^2 $. Typical time $\pt\simeq t_0/p_0$ to achieve the successful population transfer to marked states involves repeating the experiment $1/p_0$ times 
\begin{equation}
\pt=\frac{1}{\Gamma_{0}}\,\frac{\pi^2\e_0}{W}\;,\label{eq:offres}
\end{equation}
where $\Gamma_0=2\pi V^2/(W/M)$ and the first factor in r.h.s gives the typical transfer  time in absence of driver errors. Errors increase the transfer time by a large factor $\e_0/W$.

For  the maximum possible bandwidth  $W$ when nearly all states are populated, $W$ $\sim$ $\Gamma_0$$\sim$$V\sqrt{M}$,   the time of population transfer (\ref{eq:offres}) is 

\begin{equation}
\pt \sim t_{\rm G} \left(t_{\rm G} \e_0 \right)  \quad (\e_0 \gg t_{\rm G}^{-1}\sim V\sqrt{M})\;.\label{eq:tptG}
\end{equation}
As expected, when  the driver error exceeds inverse  Grover time $1/t_{\rm G}$ the performance of analogue Grover  algorithms  (\ref{eq:HG})  degrades  relative  to $t_{\rm G}$. 
 This is a direct consequence of the fact that the quantum evolution  begins from fully symmetric state which is a ground state  of the driver Hamiltonian whose energy is tuned at resonance with the marked states. In this case  the transverse field Hamiltonian driver effectively corresponds to  the projector (\ref{eq:HG}). Because the ground state is not degenerate, the resonance region is exponentially narrow ($\sim 2^{-n/2}\sqrt{M}$). This  results in the   exponential  sensitivity of the Grover algorithm performance  to the value  of driver weight. This critical behavior was studied in the work on quantum spatial search \cite{childs2004spatial} for the case of one marked state.

In contrast, in the PT protocol considered in this paper there was no need to fine-tune  the value of $\B$ other than making  it  large, $\B\gg1$. 
This happened  because   the effective coupling between the marked states described by the down-folded Hamiltonian $\scH$ (\ref{eq:IBH}) was not due to any one particular eigenstate of the driver (such as the state $\ket{S}$ for the Grover case). Instead this coupling was 
formed due to an exponentially large (in $n$)  number of  non-resonant,  virtual  transitions between the marked states and highly  exited states  of the transverse field Hamiltonian $H_D$.  This resulted in a significant  improvement in robustness for the proposed PT relative to the analogue Grover  algorithm.  

\section{Numerical simulations of population transfer}

\begin{figure}[ht]
\includegraphics[width=0.5\textwidth]
{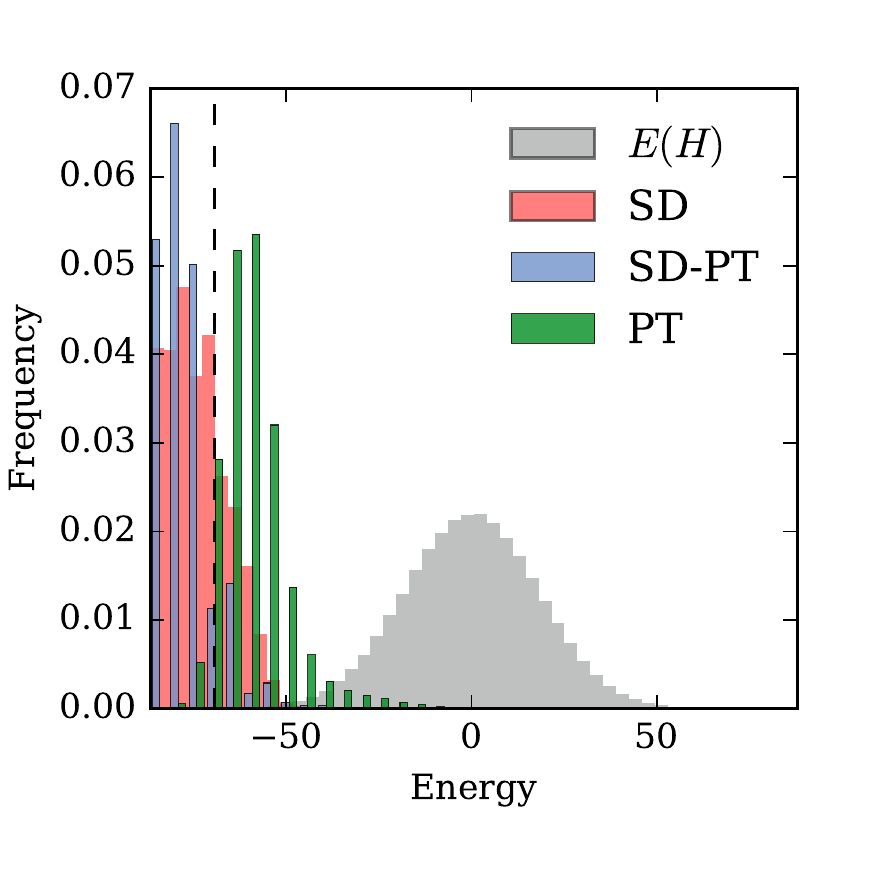}
 \includegraphics[width=0.5\textwidth]{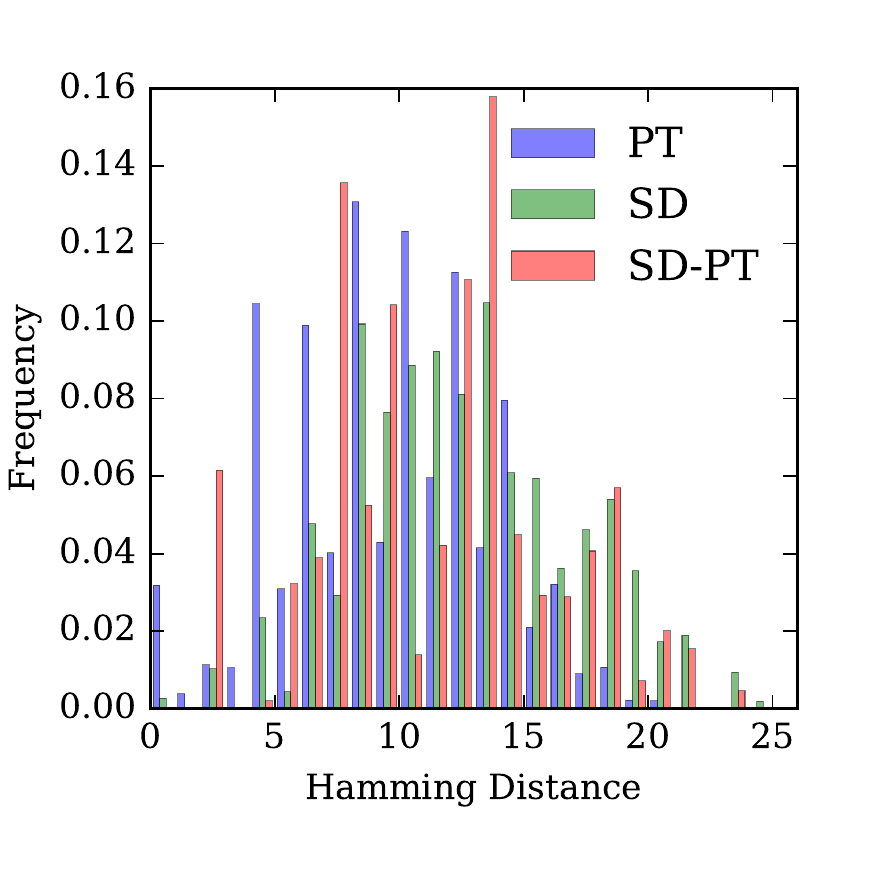}
  \caption{\textit{ Left panel} shows a histogram of normalized weights of classical energies in $H_{cl}^{(2)}$. $E(H)$ is the density of states in the original Hamiltonian spectrum. The classical steepest descent (SD) distribution shows the probability of ending up in a local minimum following a steepest descent run performed greedily by single spin flips. PT histogram shows the weight of classical energies in the output wave function  following a population transfer (PT) run relative to the initial state energy (dotted black line).  
The SD-PT run is the distribution of local minimum starting from a state measured from the PT state.  We note that the SD and SD-PT distributions are plotted as adjacent histograms to increase visibility of the data, but depicted bins are actually overlapping and not alternating.   
\textit{ Right panel} shows the distribution of Hamming distances from the initial state for states that fall within 1 standard deviation of the energy from the peak of the population transfer (PT) distribution for several methods.  
}\label{fig:PToutput}
\end{figure}

\begin{figure}[ht]
\includegraphics[width=0.5\textwidth]{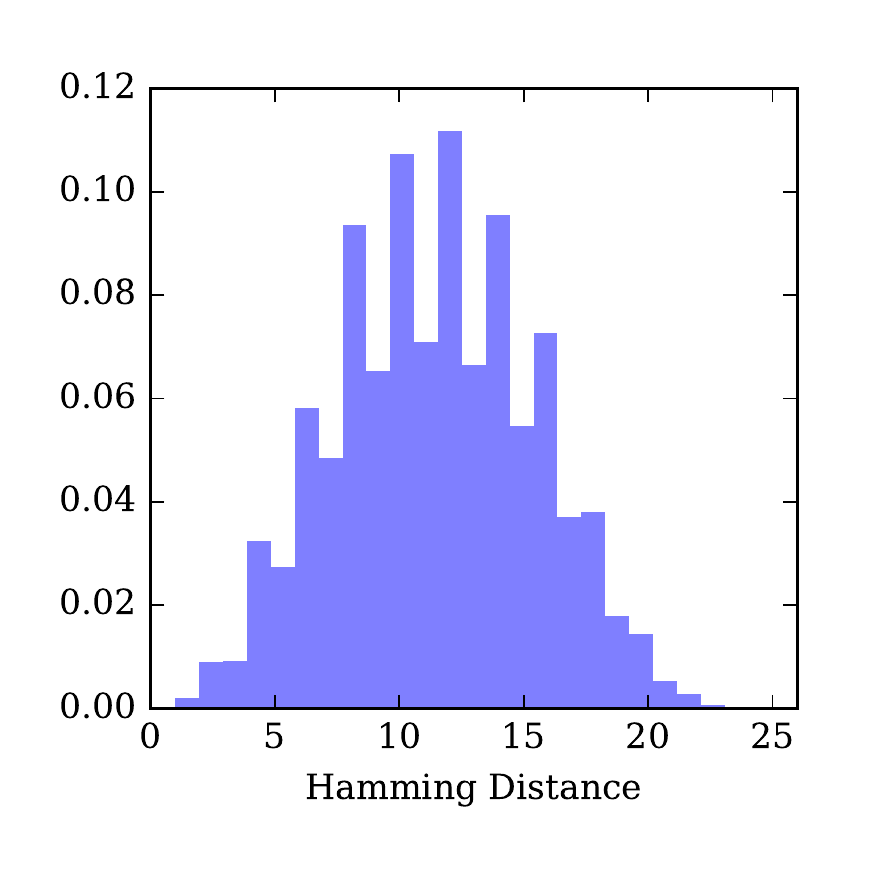}
 \includegraphics[width=0.5\textwidth]{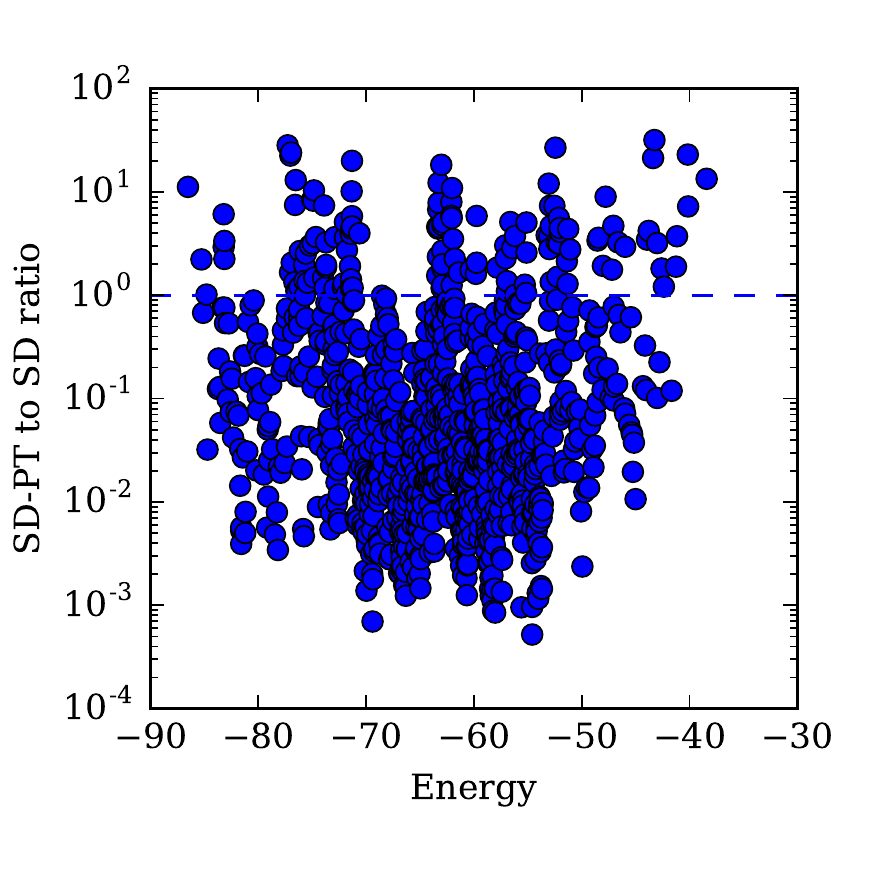}
  \caption{ \textit{ Left panel} shows the histogram of Hamming distance between all pairs of states within 1 standard deviation of the energy from the peak of the population transfer (PT) distribution. Hamming distances are weighted by their joint probability within the PT distribution.  The 0 hamming distance self contributions are excluded.  
The even-odd pattern that is observed results from the dimer structure of the Hamiltonian and the corresponding match driver. 
\textit{ Right panel} plots every local minimum with respect to single spin flips in the spectrum, which is concentrated to the low energy sector of the total Hamiltonian.  For each local minimum we depict the ratio of the probability of ending up in that local minimum when starting from the PT distribution against a uniformly random initial state.  We see that a significant fraction of states are enriched in the PT case, including the global minimum.
  }\label{fig:PTstructure}
\end{figure}

We illustrate by numerical simulations the potential of the PT subroutine for hybrid quantum-classical search and optimization algorithms. We consider a model defined by the following 2-local Hamiltonian,
\begin{equation}
\mathcal{H}=\mathcal{H}_{\cl} + \Gamma \sum_{i,j}^n  \left( 
\delta_{ij}(\left| h_i\right|+1)
\sigma_i^x +  
(\left| J_{ij}\right|+1) 
\sigma^x_i \sigma^x_j\right),\;\; \mathcal{H}_{\cl}= \sum_{i=1}^n  h_i  \sigma_i^z +\sum_{i,j}^{n} J_{ij} \sigma^z_i \sigma_j^z, \label{eq:Hnum}
\end{equation}
where $\delta_{ij}$ is the Kronecker symbol, $J_{ij}\in \left[-1,1\right]$ and $h_i \in \left[-1,1\right]$ are uniformly distributed random numbers with finite 6-bit precision. A subset of $n/2$ bonds $\left(i,j\right)$ are chosen to be dimers, non-overlapping pairs of spins with strong ferromagnetic interactions $J_{ij}=-4$. Note that we are using "matched" driver whose strength scales with the total longitudinal field acting on a given qubit.

We simulate of $n=25$ qubit system. A single bit flip steepest descent (SD) algorithm starting from all possible bistrings identifies all local minima of a given realization of the model $\mathcal{H}_{\cl}$ in Eq.~(\ref{eq:Hnum}). Optimization of simulated annealing parameters for this type of instances suggested that SD (low temperature limit) is near optimal and therefore can serve as a proxy for hardness of finding a given bitstring, see histogram in the left panel of Fig.~\ref{fig:PToutput}. Starting from a low-energy local minimum we perform a population transfer, Hamiltonian evolution with a fixed strength of the driver strength $\Gamma=0.2$ for sufficiently long time. The quantum evolution is simulated using Trotter decomposition with 300 steps. The evolution time is chosen sufficiently long for the population transfer to approximately saturate. The histogram of weights of classical energies in the output wave function, Fig.~\ref{fig:PToutput}, shows significant weight remaining in the low energy region of the spectrum in the vicinity of the initial bitstring energy.  The PT output wave function has support on bitstrings separated from the initial state by large Hamming distances, see right panel in Fig.~\ref{fig:PToutput}. Moreover, repeated sampling from the PT output produces bitstrings separated by large Hamming distances from each other, see left panel in Fig.~\ref{fig:PTstructure}. Bitstrings sampled from the PT output wave function can be used as starting states for SD which finds some of the low energy minima with higher probability than SD initialized with uniformly random bitstrings, see right panel in Fig.~\ref{fig:PTstructure}. Therefore PT protocol provides a quantum coherent pathway  between low energy states that is complementary to SD and simulated annealing and therefore PT could potentially serve as useful subroutine in hybrid algorithms such as quantum parallel tempering. 

\begin{figure}[ht]
\includegraphics[width=0.5\textwidth]
{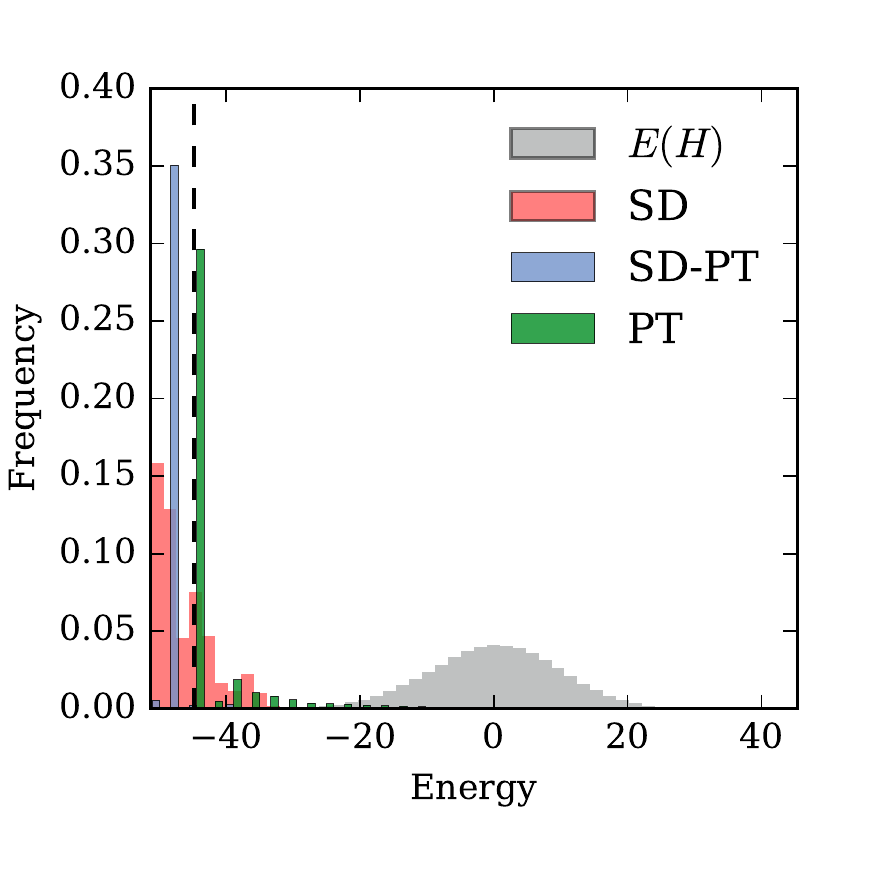}
 \includegraphics[width=0.5\textwidth]{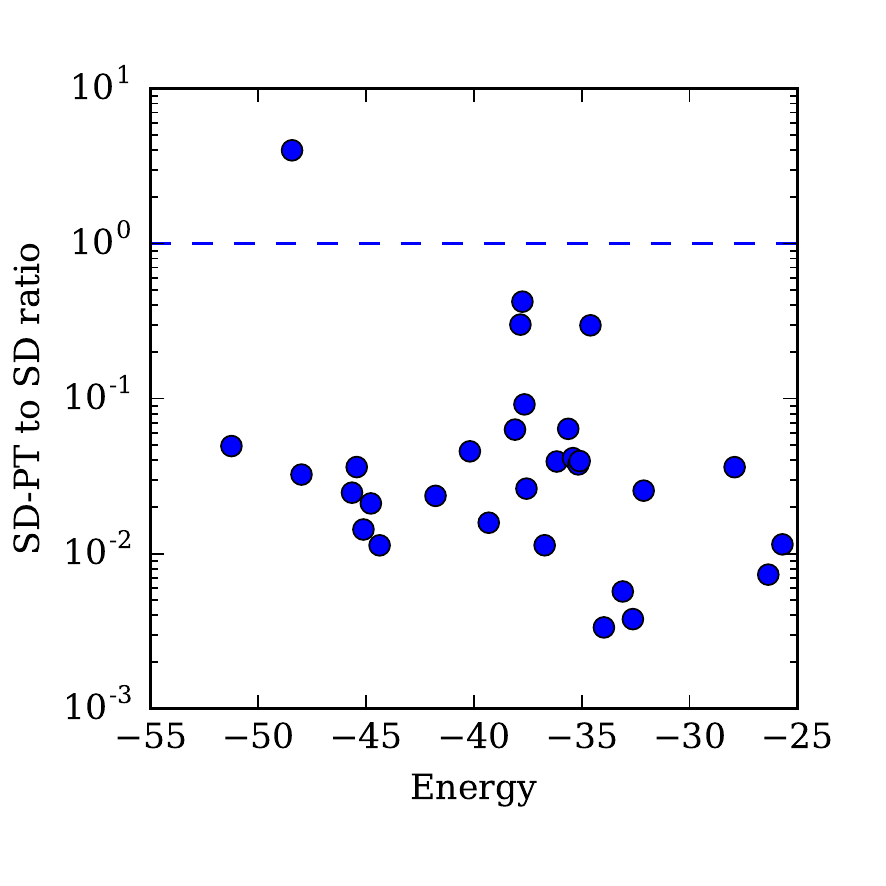}
  \caption{ For comparison we include the analog of Figs.~\ref{fig:PToutput} and~\ref{fig:PTstructure} (for $n=22$) for a model without dimers, where SD performs better.  
}\label{fig:PToutput22nodimers}
\end{figure}

\section{\label{sec:conc}Conclusion}

We analyze the computational role of coherent multiqubit tunneling that gives rise to bands of nonergodic delocalized quantum  states as a coherent pathway for population transfer (PT) between computational states with close energies.  In this regime PT cannot be efficiently simulated by QMC.

We solve this problem  using the  following quantum population transfer (PT) protocol: prepare the system in a computational state $\ket{z_j}$ with classical energy $\cE(z_j)$, then evolve it with the transverse-field quantum spin Hamiltonian. Classical energies $\cE(z)$ are encoded in the problem Hamiltonian diagonal in the basis of states $\ket{z}$. 
The key difference between PT protocol and QA~\cite{kadowaki1998quantum,farhi2001quantum,brooke1999quantum} or analogue quantum search Hamiltonians~\cite{farhi1998analog,childs2004spatial} is that the transverse field is kept constant throughout  the algorithm and  is not fine-tuned to any particular value.   At the final moment of PT  we
projectively measure in the computational basis and check if the
outcome $z$ is a ``solution'', i.e.,  $z \ne z_j$, and the energy
$\cE(z)$ is inside the window $\Delta \cE_{\cl}$.

  In this paper we  analyzed PT dynamics in  Impurity Band (IB) model with a ``bimodal'' energy function: $\cE(z)=0$  for all states  except for  $M$ ``marked''  states  $\ket{z_j}$ picked at random with energies forming a narrow band of the width $W$ separated by a large gap $\cO(n)$ from the rest of the states.  This landscape is  similar to that in analogue Grover   search   \cite{farhi1998analog,farhi2000quantum} with multiple target states and a distribution of oracle values for the targets. The best known classical algorithm for finding another marked state has cost $O(2^n/M)$. 

The PT dynamics  is described by the down-folded
$M\times M$ Hamiltonian $\scH$ that is dense in the space of the marked states $\ket{z_j}$. The distribution of matrix elements $\scH_{ij}$  has a heavy tail decaying as a cubic power for $V(d)\gg V_{\rm typ}$. This is a remarkable result of the competition between the very steep decay of the off-diagonal tunneling matrix element with the Hamming distance $d$, and the steep increase in the number of marked states $M_d\propto\binom{n}{d}$ at distance $d$. We emphasize that such polynomial tail in the distribution of matrix elements is only possible either in infinite dimension or in presence of long-range interactions (e.g, dipolar glass). 

The dispersion of the diagonal elements  $\scH_{jj}=\cE(z_j)$ is expected to be large, $W\sim V_{\rm typ} M^{\gamma/2} \gg V_{\rm typ} $ with $\gamma\in\left[1, 2\right]$.  In the range $1<\gamma<2$ there  exist  minibands of non-ergodic delocalized eigenstates of $\scH$. Their  width is proportional to $1/\pt \ll W$.   Each miniband associated with a support set $\scS$ over the marked states.   

The distribution of miniband widths $\Gamma$  obeys alpha-stable Levi law with tail index 1. The typical value of $\Gamma$ and its  characteristic variance  exceeds the typical matrix element of $\scH$ by  a factor $\Omega^{1/2}$ where $\Omega=(M V_{\rm typ}/W)^2$ is a size of the support set in a typical miniband.

We demonstrate that quantum PT finds another state within a target window of energies $\Omega$ in time $\pt \propto 2^{n/2} \Omega^{-1/2} \exp( n/(2B_\perp^2))$. The scaling exponent of $\pt$ with $n$ differs from that in Grover's algorithm by a factor $\propto B_\perp^{-2}$, which can be made small with large transverse fields $n \gg B_\perp^2 \gg 1$. 

Crucial distinctions between this case and the Hamiltonian in the  analogue version of Grover's algorithm \cite{farhi1998analog}  for the case of multiple target states are the
non-integrability of our model, and the delocalized nature of the
eigenstates within the energy band $W$.  Furthermore, analogue 
Grover's algorithm for multiple targets is  exponentially sensitive in $n$ to the
weight of the driver Hamiltonian,  and cannot be initialized with a
computational basis state. 

The quantum spin Hamiltonian in  (1)  can be described using n-body, infinite-range interactions. However, it shares key properties with the infinite range spin-glass models involving only p-body interactions  at finite values of p >2 that could be implemented on a quantum computer with polynomial in n resources. The evidence of non-ergodic extended states were recently uncover numerically in the low-energy part of the spectrum of the quantum transverse field p-spin model~\cite{BaldwinLaumannPSpin2017}. 

Similar to the model  (1) the low-energy part of the spectrum of transverse field p-spin model is characterized by the proliferation of statistically independent deep local minima  separated by large, O(n), number of spin flips. Model  of this type are characterized by a RSB-1 type of spin glass transition and were studied in the context of   quantum annealing~\cite{Jorg}. They represent perhaps the next step to study PT algorithms.

Finally, we analyzed numerically the PT initiated at a low energy local minimum of a 2-local spin glass model and observed that sampling PT output together with subsequent application of classical steepest descent allows exploring the energy landscape in a way complimentary to steepest descent and simulated annealing. This suggest possible use of PT as a subroutine for hybrid quantum-classical algorithms for search and optimization such as quantum paralleled tempering.





\end{document}